\documentstyle[eqsecnum,aps,preprint,epsfig]{revtex}

\def\laq{\ \raise 0.4ex\hbox{$<$}\kern -0.8em\lower 0.62
ex\hbox{$\sim$}\ }
\def\gaq{\ \raise 0.4ex\hbox{$>$}\kern -0.7em\lower 0.62
ex\hbox{$\sim$}\ }
\tightenlines
\def\half{\hbox{\magstep{-1}$\frac{1}{2}$}}
\def\third{\hbox{\magstep{-1}$\frac{1}{3}$}}
\def\twothd{\hbox{\magstep{-1}$\frac{2}{3}$}}
\def\quarter{\hbox{\magstep{-1}$\frac{1}{4}$}}
\def\sixth{\hbox{\magstep{-1}$\frac{1}{6}$}}
\def\twelve{\hbox{\magstep{-1}$\frac{1}{12}$}}
\def\thrhaf{\hbox{\magstep{-1}$\frac{3}{2}$}}
\newcommand{\diff}[1]{{d#1 \over d\tau}}

\def\PRD{{\em Phys. Rev.} D}

\begin{document}
\draft
\preprint{\vbox{\baselineskip=12pt
\rightline{CERN-TH/98-24}
\vskip0.2truecm
\rightline{UPRF-98-1}
\vskip0.2truecm
\rightline{gr-qc/9802001}}}
\title{A numerical simulation of pre-big bang cosmology}
\author{J. Maharana$^1$\footnote{Permanent address: Institute of
Physics, Bhubaneswar 751 005, India.},
E. Onofri$^2$ ,
 and G. Veneziano$^1$}
\address{$^1$Theory Division, CERN, CH-1211 Geneva 23, Switzerland}
\vskip 2 cm
\address{$^2$ Dipartimento di Fisica, Universit\`a di Parma and
INFN, Gruppo Collegato
di Parma,\\
  43100 Parma, Italy}
\vskip 2. cm
\maketitle
\begin{abstract}
We analyse numerically the onset of pre-big bang inflation in an
inhomogeneous, spherically symmetric Universe.  Adding a small
dilatonic perturbation to a trivial (Milne) background, we find that
suitable regions of space undergo dilaton-driven inflation and quickly
become spatially flat ($\Omega \rightarrow 1$).  Numerical
calculations are pushed close enough to the big bang singularity to
allow cross checks against previously proposed analytic asymptotic
solutions.

\end{abstract}

\vskip 5 cm

\noindent CERN-TH/98-24 \\
January  1998
\newpage

\global\firstfigfalse   %  Avoid \newpage and headline ``FIGURES'' 

%%%%%%%%%%%%%%%%%%%%%%%%%%%%%%%%%%%%%%%%
\section{INTRODUCTION}
%%%%%%%%%%%%%%%%%%%%%%%%%%%%%%%%%%%%%%%%

Increasing attention has recently been devoted to a possible
alternative to the standard inflationary paradigm, the so-called
pre-big bang (PBB) scenario \cite{GV},\cite{MGV}, \cite{MG}. While
referring the interested reader to \cite{reviews} for recent reviews
of the subject, we will start by just recalling the essential points
needed to put the present work in the correct perspective.

The basic postulate of PBB cosmology is, at first sight, a shocking
one: our Universe would have originated from an ``anti-big bang"
state, which was essentially empty, cold, flat, and decoupled. The
claim is that, from such innocent-looking initial conditions, a rich
Universe can originate thanks to two distinct mechanisms:
\begin{itemize}
\item[\sl i)] A {\sl classical} gravitational instability amplifies
tiny initial perturbations, inevitably pushing the Universe towards a
singularity in the future (to be later identified with the standard
big bang) through a phase of accelerated expansion (inflation) and
accelerated growth of the coupling.  We refer to this phase as
dilaton-driven inflation (DDI).
\item[\sl ii)] {\sl Quantum} fluctuations are amplified during DDI
according to the phenomenon by which perturbations freeze out as their
wavelength is stretched beyond the Hubble radius. This is how we are
able to produce a {\sl hot} big bang at the end of DDI.
\end{itemize}

Obviously, in order for the whole scenario to be viable, a mechanism
has to be conceived to produce the exit from the dilaton-driven phase
to the radiation-dominated phase of standard cosmology. This is the
so-called exit problem \cite{exit}, on which we will have nothing new
to say in this paper.  Rather, our attention will be focused on the
pre-big bang classical epoch, characterized by small couplings and
small curvatures (in string units). The big simplification that occurs
in this epoch is that the field equations are basically known since
they follow from the low-energy tree-level effective action of string
theory.

One would like to discuss the most general solution to the field
equations and check under which conditions PBB inflation takes place
and is sufficiently efficient to produce something like the patch of
the Universe presently observable to us.  There have been claims
\cite{TW} that this calls for highly fine--tuned initial
conditions. On the other hand, arguments given in \cite{GV1},
\cite{BMUV1} have suggested the following interesting
possibility/conjecture: the generic initial Universe that is able to
give dilaton--driven inflation is one which, in the asymptotic past,
converges to the Milne metric with a constant dilaton. Such an initial
state does not look generic at first: it is so, however, in a
technical sense, i.e. in that the general solution that develops PBB
inflation is claimed to depend on as many arbitrary functions of space
as the most general solution does.  If confirmed, this would mean that
PBB inflation covers a non-vanishing fraction of the total phase space
(the space of all classical solutions) of string theory.

The main purpose of this paper is to provide a non--trivial check of
the above conjecture.
%Because of present computer limitations %(???Enrico please check!),
%we shall not be able to address the full
We shall not address the full inhomogeneous problem, for the moment,
which would require a much stronger numerical effort; we will limit
our attention, instead, to the case of a spherically symmetric
Universe. We thus consider the problem of how a small spherically
symmetric lump (alternatively a shell) of energy affects the otherwise
trivial evolution of Milne's metric.  We mention that other
interesting questions in spherically symmetric pre-big bang cosmology
have been recently addressed by Barrow and Kunze \cite{Barrow}.

In Section 2 we define our choice of gauge for the metric and write
down the field equations.  As usual these break up in two sets:
constraints, containing only first time-derivatives, and
(constraint-preserving) evolution equations. Luckily, we are able to
solve the constraints in closed form and to reduce the equations to
four first-order partial differential equations (PDEs) in one time 
and one space.  In Section 3 we apply the general gradient expansion
method \cite{GE} to our particular case and construct analytic
asymptotic solutions near the singularity.  In Section 4 we outline
our technique for numerically solving the equations and discuss some
subtleties needed to avoid possible singularities at $r=0$.  The
numerical results, both for a spherical lump and for a shell, are
reported in Section 5, where comparison and matching with the
analytic asymptotic formulae derived in Section 3 are also made.
Finally, in Section 6, we interpret our results in the string frame
and discuss  the relevance of our work to the fine-tuning question
raised in \cite{TW}.

%%%%%%%%%%%%%%%%%%%%%%%%%%%%%%%%%%%%%%%%
\section {Field equations and elimination of the constraints}\label{fieldeq}
%%%%%%%%%%%%%%%%%%%%%%%%%%%%%%%%%%%%%%%%

In this paper we shall limit our attention to the graviton--dilaton
system in four space--time dimensions. Other fields, such as the 
axion
and various moduli, are taken to be frozen. The low-energy 
tree-level
effective action in the physical (string) frame (whose metric is
denoted by $G$) reads \cite{Lovelace}, \cite{Callan}:
\begin{equation}
\hbar^{-1}~S_{\rm eff} ={1 \over 2~\lambda_s^2}~\int d^4x\, \sqrt {-G}
\, e^{-\phi} \, \left( R(G) + G^{\mu\nu}\, \partial_{\mu} \phi\,
\partial_{\nu} \phi\right)
\end{equation}
where $\lambda_s$, the string-length parameter, will play no role in
the classical regime discussed here. We shall come back to its role in
the later evolution in Section 6.  The Einstein frame, which turns out
to be somewhat more convenient for solving the equations, is related
to the string frame by $8 \pi G_{\mu\nu} = e^{\phi - \phi_0}
g_{\mu\nu}$, with $\phi_0$ denoting the present value of the dilaton.
In that frame, defining $ 8\pi l_P^2 = \rm{exp}(\phi_0)
\lambda_s^2$, the action becomes \cite{Callan}:
\begin{equation}
\hbar^{-1}~S_{\rm eff} =
{1 \over 16 \pi l_P^2} \int d^4x \sqrt{-g} \left( R(g) -
\half g^{\mu\nu}\, \partial_{\mu}\phi\, \partial_{\nu}\phi \right)
\end{equation}
and leads to Einstein's equation (after eliminating its trace):
\begin{equation}
R_{\mu \nu} = \half \partial _\mu \phi \,\partial_\nu \phi
\end{equation}
and to the dilaton evolution equation:
\begin{equation}
\partial _\mu(\sqrt
{-g}g^{\mu \nu} \partial_\nu \phi) = 0 \; .
\end{equation}
For spherically symmetric cosmological solutions, $\phi$ depends just
on time $t$ and on a radial coordinate $\xi$. In the synchronous gauge
we can use spatial coordinates such that the metric takes the form
(see e.g. \cite{Weinberg}):
\begin{equation}
 ds^2 = -dt^2 + e^{2\alpha
(\xi,t)}\,d\xi^2+ e^{2\beta (\xi,t)} \, d\omega^2\; .
\end{equation}
where $d\omega^2=d\theta ^2 +\sin^2 \theta\,d\varphi^2$. 
Let us denote by a dot (a prime) differentiation with respect to
$t$ ($\xi$),  introduce
\begin{equation}
 \dot{g}/g \equiv \chi = 2 \dot{\alpha} + 4 \dot{\beta}~,
~\Delta = 2\left( \dot{\alpha} - \dot{\beta} \right) \; ,
\end{equation}
and recall the expression for the three-curvature $^{(3)}R$ on a
constant $t$ hypersurface
\begin{equation}
^{(3)}R =  2 e^{- 2\beta} - 2 e^{- 2\alpha} \left(2 \beta'' +
3 \beta'^2 - 2 \alpha' \beta' \right)\; .
\end{equation}
In terms of these, the full set of field equations consists of

\noindent {\sl i)} the Hamiltonian and momentum constraints:
\begin{eqnarray}
\third(\chi ^2 - \Delta ^2) + 2 ~{^{(3)}R} &=& {\dot{ \phi}}^2
+ e^{- 2\alpha}\,\phi'^2
 \nonumber \\
  2 \beta' \Delta - \twothd
 (\chi - \Delta)'  &=& {\phi}'\dot{\phi}\; ,
\label{constr}
\end{eqnarray}

\noindent
{\sl ii)} the evolution equations for the metric
\begin{eqnarray}
{\dot {\chi}} &=& - \sixth \left( \chi ^2 + 2 \Delta ^2 \right) -
{\dot {\phi}} ^2 \nonumber \\
\dot {\Delta}+ \half \chi \Delta &=&
	e^{- 2\alpha}\,\phi'^2 + 2 e^{- 2\beta} + 2 e^{- 2\alpha}
	\left(\beta''- \alpha' \beta' \right) \; ,
\label{evol}
\end{eqnarray}
and, finally,
\par\noindent
{\sl iii)} the equation for the dilaton
\begin{eqnarray}
\ddot {\phi}+ \half \chi \dot {\phi} =e^{- 2\alpha} \left({\phi}''
+(2 \beta' - \alpha'){\phi}'\right)\; .
\label{dilevol}
\end{eqnarray}

This last equation can be shown to be a consequence of the previous
ones and we shall therefore ignore it. One can also show that the
constraints are conserved in time, thanks to the explicit form of 
the evolution equations.  The Hamiltonian constraint can be written in
terms of the well-known $\Omega$ parameter (the fraction of critical
energy density) as:
\begin{equation}
\Omega \equiv \rho/\rho_{cr} = 3\,{\dot{
\phi}^2 + e^{-2\alpha}\,\phi'^2 \over \chi^2 - \Delta^2 } = 1 +
6\, {{^{(3)}R}\over \chi^2 - \Delta^2}\;.
\end{equation}
A nice simplification, which occurs in the spherically symmetric case,
is that the constraints (\ref{constr}) can be solved explicitly in
terms of $\dot{\phi}$ and $\phi'$. Given the quadratic character of
the constraints, this leads to a four-fold ambiguity, reminiscent of
the two-fold ambiguity of the homogeneous case, which we resolve by
imposing (in accordance with the PBB postulate) a monotonic behaviour
for $\phi$ at least from some time on. We then get:
\begin{eqnarray}
\dot {\phi} &=& \sqrt {A^2 + B} + \sqrt {A^2 - B} \ge 0 \nonumber \\
\phi' &=& e^{\alpha} \left( \sqrt {A^2 + B} - \sqrt {A^2 - B}\right)\;,
\label{solconstr}
\end{eqnarray}
where the auxiliary quantities $A$ and $B$ are given by
\begin{eqnarray}
 A^2 &=& \twelve (\chi ^2 - \Delta ^2) + e^{-2\beta}- e^{-2\alpha}(
2\beta '' + 3\beta'^2 - 2\alpha ' \beta ') \nonumber \\
B &=& e^{-\alpha}(\beta ' \Delta - \third (\chi -\Delta)')\;.
\label{AB}
\end{eqnarray}

In conclusion, the final system of evolution equations contains only
the metric and can be written in first-order form as:
\begin{eqnarray}
{\dot {\alpha}} &=& \sixth \left( \chi + 2 \Delta \right) \nonumber \\
{\dot{\beta}} &=& \sixth \left( \chi - \Delta \right) \nonumber \\
{\dot{\chi}} &=& - \sixth \chi ^2 - \third \Delta ^2 - 2(A^2 +
	\sqrt{A^4-B^2}) \nonumber \\
{\dot {\Delta}} &=& - \half \chi \Delta +
2\,e^{-2\beta} + 2\,e^{-2\alpha} (\beta '' - \alpha ' \beta ') +
	2(A^2 - \sqrt {A^4 - B^2})\;,
\label{finalsystem}
\end{eqnarray}
with $A,B$ given in Eq.\ (\ref{AB}).  By using the above relations and
the expressions for $\dot {\phi}$ and $\phi '$ given in
Eq.\ (\ref{solconstr}), it is explicitly checked that the
``integrability condition" $(\dot \phi )' = (\phi ')^{\bf .}$ holds.

An essentially trivial solution to all equations is given by the
so-called Milne metric, known in the literature \cite{Milne} because
it represents the very late time behaviour (with $\Omega \rightarrow
0$) of all subcritical Universes (with $\Omega < 1$). In our gauge,
Milne's metric corresponds to:
\begin{equation}
\alpha = \ln (\pm t) \; , \; \beta = \alpha + \ln \sinh \xi \;,
\; \phi = \rm{constant}\;.
\end{equation}
It is well known that Milne's metric can be brought to Minkowski's
form through a coordinate transformation that maps the whole of
Milne's space-time into the (interior of the) forward or backward
light cone (depending on the sign of $t$) of Minkowski space-time.
The conjecture advanced in \cite{BMUV1} is that pre-big bang inflation
finds its generic origin in open cosmologies that approach Milne as
$t\rightarrow -\infty$.  This is why we will choose initial data (at
some finite negative $t$) very close to Milne's Universe, by inserting
some lump of energy/momentum through a non-trivial  dilaton.

Generic initial data in the spherically symmetric case are known
\cite{LL} to depend on two arbitrary functions of the radial
coordinate $\xi$.  It is instructive to see how these two arbitrary
functions appear in our initial data. Apparently, in order to select a
definite solution to Eqs.\ (\ref{finalsystem}), one has to provide 
four functions of $\xi$ on a Cauchy hypersurface, i.e.  $\alpha,
\beta, \chi$, and $\Delta$ at some initial time. However, two of these
can  be eliminated by the two residual gauge transformations
\cite{LL} that keep us inside the synchronous gauge.  We may thus take
physically distinct initial data to be given by $A$ and $B$ of
Eq.\ (\ref{AB}). Appropriate combinations of $A$ and $B$ represent 
the initial energy and pressure density of (dilatonic) matter.
Equivalently, we may observe that the general spherically symmetric
solution of the dilaton evolution equation Eq.\ (\ref{dilevol}) is
given (in the case of Milne's metric) by \cite{Tanaka}
\begin{equation}
\phi = {e^{\tau} \over \sinh \xi}~\int_{-\infty}^{\infty} dp~b_p~e^{ i p \tau}
 \left( { \sin p \xi \over p }\right) ~+~ \rm{constant} \; , ~ \tau
\equiv - \ln (-t/T_0)
\end{equation}
and thus depends on two real functions of $p$ (since $b_p$ is
complex).

This latter observation allows us to make a more general remark:
string vacua are usually identified with two-dimensional conformal
field theories (CFTs) resulting from a set of $\beta$-function
equations \cite{Lovelace}, \cite{Callan}.  However, within a
particular CFT, we can construct marginal (i.e. $(1,1)$) vertex
operators representing physical excitations (particles) propagating
in the given background.  A physical dilaton vertex operator depends
upon two (real) functions of space, while a physical graviton vertex
depends on four.  It is tempting to identify these two (resp. six)
degrees of freedom with the classical moduli of our spherically
symmetric (resp. generic) solutions, since we can indeed expect
 a correspondence between marginal operators and the structure of
moduli space in the neighborhood of a CFT.

In giving initial data, two precautions have to be taken: i) we should
verify that, at least initially, $A^2 > |B|$, so that all the square
roots appearing in Eqs.\ (\ref{finalsystem}) are real.  Consistency
requires that this constraint be maintained through the evolution; ii)
there is an apparent singularity at $\xi=0$ in $\beta_{Milne}$ and in
its derivatives.  Of course the singularity is perfectly canceled for
Milne's metric in all equations, but one may encounter numerical
problems --or even genuine physical ones-- if one is not careful about
the way to perturb Milne. We believe that the correct way to avoid
curvature singularities and to achieve a smooth algorithm is to insist
on an ansatz of the form:
\begin{equation}
\alpha =
\alpha(\xi^2, t)\; , \; \beta = \alpha + \ln \sinh \xi + \xi^2
\delta(\xi^2, t)\;,
\label{condition}
\end{equation}
giving
\begin{equation}
\chi = 6~\dot{\alpha} + 4~\xi^2
\dot{\delta} \; , \; \tilde{\Delta} \equiv \xi^{-2} \Delta = 
- 2 \dot{\delta}\;.
\label{smooth}
\end{equation}
It can be checked that, rewriting all equations in terms of $\alpha,
\delta, \chi, \tilde{\Delta}$, and their derivatives with respect to
$\xi^2$, all singularities at $\xi =0$ are automatically removed and
that the structure of the ansatz is maintained during the evolution.
A possible way to construct sets of ``good'' initial data along these
lines is discussed in the appendix.

%%%%%%%%%%%%%%%%%%%%%%%%%%%%%%%%%%%%%%%%
\section {Analytic asymptotic solutions near the singularity}\label{asympt}
%%%%%%%%%%%%%%%%%%%%%%%%%%%%%%%%%%%%%%%%
In this section we use, as in \cite{GV1},\cite{BMUV1}, the gradient
 expansion technique \cite{GE} to construct analytic asymptotic
 solutions to our spherically symmetric field equations. This exercise
 is important in several respects.  Firstly, it illustrates the
 gradient expansion technique in a situation where the momentum
 constraints can be explicitly solved. Secondly, it will allow a very
 non-trivial check of the numerical method. Finally, it allows the
 numerical solutions to be continued in a region where the computation
 undergoes a critical slowdown because of the singularity. Hence, in
 the problem at hand, analytic and numerical methods very nicely
 complement each other.

Assuming that spatial gradients become subleading near the
singularity,we can simplify Eqs.\ (\ref{constr}), (\ref{evol}) to the
following form:
\begin{eqnarray}
\dot{\phi}^2 = \third ( \chi^2 - \Delta^2)~~~ &,& ~~~
 {\phi}' =  \left(2 \beta' \Delta - \twothd (\chi - 
\Delta)'\right)\big/
\dot\phi\;, \nonumber \\
\dot{\chi} + \half \chi^2 = 0 ~~~ &,& ~~~ 
	\dot{\Delta} + \half \chi\Delta = 0\;.
\end{eqnarray}
Following the procedure of Refs. \cite{GV1}, \cite{BMUV1} we first
solve the evolution equations and obtain:
\begin{equation}
\chi = {2 \over t - t_0(\xi) } \; ,\; \Delta = {2 \lambda(\xi)
\over  t - t_0(\xi)}\;.
\label{asy1}
\end{equation}
The Hamiltonian constraint can be  easily solved. Choosing the
branch corresponding
to a growing dilaton we obtain:
\begin{equation}
\phi(\xi, t) = \phi_0(\xi) - {2 \over \sqrt{3}} \sqrt{1 - \lambda^2}
\ln \left({t / t_0(\xi)} -1\right) \; ,\; t < t_0\;.
\end{equation}
We now integrate once more the evolution equations to obtain the
metric:
\begin{eqnarray}
\alpha &=& \gamma(\xi) +  \third \left( 1 + 2 \lambda(\xi)\right)\,
\ln \left({t / t_0(\xi)} -1\right)\;, \nonumber \\
\beta &=& \delta(\xi) +  \third \left( 1 - \lambda(\xi) \right)\,
\ln \left({t / t_0(\xi)} -1\right)\;.
\label{asy2}
\end{eqnarray}

Finally, we solve the momentum constraint. Most terms automatically
match and we get the single condition:
\begin{equation}
\phi_0' = - {2  \sqrt{3} \over \sqrt{1 - \lambda^2}}
\left(\lambda'/3 + \lambda \delta' \right)\;.
\end{equation}
The general solution thus appears to depend upon four functions of
space, i.e. $\lambda, \gamma, \delta$, and $t_0$, giving the
line-element:
\begin{equation}
ds^2 = - dt^2 + \left(t/t_0 -1 \right)^{2/3(1-\lambda)}
\left[\left(t/t_0 - 1\right)^{2 \lambda}\,
e^{2\gamma} \,d\xi^2 + \,
e^{2\delta} d\omega^2 \right]  \nonumber
\end{equation}
Obviously, we can reabsorb $\gamma$ in a redefinition of the $\xi$
coordinate, $dr = e^{\gamma} d\xi$. Similarly, the choice of
equal-time slices allows us to remove the space dependence of $t_0$.
We are thus left, as in the previous section, with just two physically
meaningful functions of $\xi$ for the characterization of our
dynamical system, as should be the case for the general spherically
symmetric solution \cite{LL}.

Actually, the situation is a bit more subtle as far as the $\xi$
dependence of $t_0$ is concerned. Inserting the above asymptotic
solution back into the exact equations, one finds that spatial
derivatives are only subleading for sufficiently small values of
$t_0'(\xi)$.  In other words, at least in the Einstein frame, one {\sl
has} to choose appropriate time slices so that $t_0$ is constant and
only then can one neglect spatial derivatives. Constructing these
privileged time slices is not a simple problem and therefore, in 
this paper, we will only check (see Section 5) the asymptotic formulae
in the vicinity of the minimum of $t_0$ (i.e. near the point where the
singularity is first reached).  In Section 6 we will argue that an
alternative solution to this problem consists in going over to the
synchronous gauge in the string frame.  In this case, spatial
derivatives (at fixed string-frame time) turn out to be always
subleading near the singularity. This suggests that it would be
desirable to numerically solve the equations directly in the string
frame.

%%%%%%%%%%%%%%%%%%%%%%%%%%%%%%%%%%%%%%%%
\section{The numerical approach}\label{numerical}
%%%%%%%%%%%%%%%%%%%%%%%%%%%%%%%%%%%%%%%%
We shall now briefly describe the essential aspects of the numerical
algorithm and its practical implementation.  The system given by
Eqs.\ (\ref{finalsystem}) is similar to a Hamiltonian one with an
infinite number of degrees of freedom. Its numerical implementation
must necessarily introduce a limit on the number of degrees of
freedom, which can be done in several ways. Our present choice is to
introduce a finite  grid in the variable $\xi$ and to define the
derivatives with respect to $\xi$ by a spectral method, based on the
Fourier transform, which  allows us to reach a high
precision. Alternative techniques, such as using  a symbolic language
or applying finite elements techniques, will be considered in the
near future. In our approach, a special care must be devoted to
boundary conditions, since the Fourier transform preferably works in
periodic boundary conditions, which are not adequate to our
problem. As we explain later on, this problem is solved by letting
$\xi$ extend to a symmetric interval and by continuing the fields to
the negative $\xi$-axis.  Time evolution is generated via a standard
integration routine with variable step and local error control.

To prepare the equations to be implemented in numerical terms, it is
convenient to introduce new fields, which describe the metric as a
perturbation from a Milne background:

\begin{eqnarray}
\alpha_s &=& \alpha + \tau\nonumber\\
\beta_s &=& \beta + \tau -\ln\sinh\xi\nonumber\\
\chi_s &=& e^{-\tau}\,\chi\nonumber\\
\Delta_s &=& e^{-\tau}\,\Delta
\end{eqnarray}
where $\tau$ is defined by $t = - \exp(-\tau)$. The advantage of
working in the Milne background is that the shifted fields, unlike the
background, vanish at $\xi=\infty$ and we can thus introduce a finite
volume cutoff. The boundary conditions at $\xi=0$ are crucial; as  we
have already noticed, regularity at $\xi=0$ is achieved if we  assume
Eq.\ (\ref{condition}), in the equivalent form
\begin{eqnarray}
\beta_s(\xi,\tau) &=& \alpha_s(\xi,\tau) + \xi\,
\eta(\xi,\tau)
\nonumber\\
\Delta_s(\xi,\tau) &=& \xi\, \Sigma(\xi,\tau)
\label{shifted}
\end{eqnarray}
with boundary conditions at $\xi=0$
\begin{eqnarray}
\eta(0,\tau)&=&\Sigma(0,\tau)=0 \,,\\
\alpha_s(0,\tau)'&=& \chi_s(0,\tau)'=0\;.
\end{eqnarray}
In order to satisfy these boundary conditions and still apply the 
Fourier
transform to compute field derivatives, we continue the fields to a
symmetric interval $-L < \xi < +L$; we enforce the correct boundary
conditions by extending $\alpha_s, \chi_s$ {\sl to symmetric
functions} and $\eta, \Sigma$ {\sl to antisymmetric ones}. In terms
of the new fields, the system (\ref{finalsystem}) becomes
\begin{eqnarray}
\diff{\alpha_s} &=& \sixth(\chi_s+6+2\,\xi\Sigma)\nonumber\\
\diff{\eta} &=& -\half\Sigma\nonumber\\
\diff{\chi_s} &=& -\chi_s -\sixth\chi_s^2 - \third \Delta_s^2 -
\Lambda_+\nonumber\\
\diff{\Sigma} &=& -\Sigma - \half \chi_s\Sigma +
2\,\xi^{-1}\,e^{-2\alpha_s} \,
\left( {e^{-2\xi\eta} - 1 \over \sinh^2\xi} +
\beta_s'' - \alpha_s'\,
(\beta_s' + \coth\xi)\right) +
\xi^{-1}\,\Lambda_- ,
\end{eqnarray}
where
\begin{eqnarray}
\Lambda_\pm &=& 2\,\left(A_s^2 \pm 
\sqrt{A_s^4-B_s^2}\right)\nonumber\\
A_s^2 &=& \twelve \left(\chi_s^2-
\Delta_s^2\right) + \nonumber\\
&& e^{-2\alpha_s} \left(
{e^{-2\xi\eta_s} - 1 \over \sinh^2\xi} -3 - 2 \beta_s''
-3\beta_s^{'2} -2(3\beta_s'-\alpha_s')\,\coth\xi +
2\alpha_s'\beta_s'\right)\nonumber\\
B_s &=& e^{-\alpha_s} \,\left(\Delta_s\,(\beta_s'+\coth\xi)
-\third (\chi_s'-\Delta_s') \right)\;.
\label{ABs}
\end{eqnarray}
and Eq.\ (\ref{shifted}) is understood.

Notice that the system is identically satisfied by taking
$\alpha_s=\eta=\Sigma=0$ and $\chi_s=-6$, which represents Milne's
background. The initial conditions we want to examine are given by
Eq.\ (\ref{deformation}) (with $\zeta=0$), which in terms of the
shifted fields reads as follows:
\begin{eqnarray}
\alpha_s(\xi,0) &=& \quarter\ln\left(1 - {\mu \over
\cosh[\varepsilon(\xi-\xi_0)]}\right)\,+\,(\xi_0 \longrightarrow
-\xi_0) \nonumber\\ \chi_s(\xi,0) &=&
-6\,\exp\{-2\alpha_s(\xi,0)\}\label{ansatz}\\ \eta(\xi,0) &=&
\Sigma(\xi,0) = 0\nonumber , \end{eqnarray} where $\mu$ is a suitable
constant.  We have implemented this system of partial differential
equations in {\sc matlab$^{\cite{mat}}$}, which offers very efficient
built-in routines for integrating ordinary differential equations, a
fast Fourier transform and an integrated graphic
environment.\footnote{A copy of the code is available at the {\sc url}
{\tt http://www.fis.unipr.it/$\scriptstyle{\sim}$onofri}.} The only tricky point in
the numerical treatment of these equations refers to the delicate
cancelation mechanism that makes the solutions regular at $\xi=0$ in
spite of the apparent singular nature of the terms involving
$\coth\xi$ and $\sinh^{-2}\xi$. This cancelation, which takes place
in exact arithmetic, is spoiled by finite precision arithmetic and
makes the solution singular after a short time evolution. The remedy
that we adopt consists in enforcing regularity near the origin by a
polynomial interpolation and a Fourier filtering, which truncates high
frequency components above a certain cutoff. A consistency check is
given by solving the equations with various initial {\sl ansatze}
corresponding to $\xi_0=0$ or $\xi_0>0$. The former case represents a
spherically symmetric perturbation concentrated at the origin, while
the latter spreads out over a spherical shell and naturally avoids the
singularity. A similar behaviour is indeed observed in both cases with
fields $\alpha$ and $\chi$ blowing up as $\Omega$ approaches 1.

%%%%%%%%%%%%%%%%%%%%%%%%%%%%%%%%%%%%%%%%%
\section{Numerical results}
\begin{figure}[h]
\begin{center}
{\mbox{\epsfig{figure=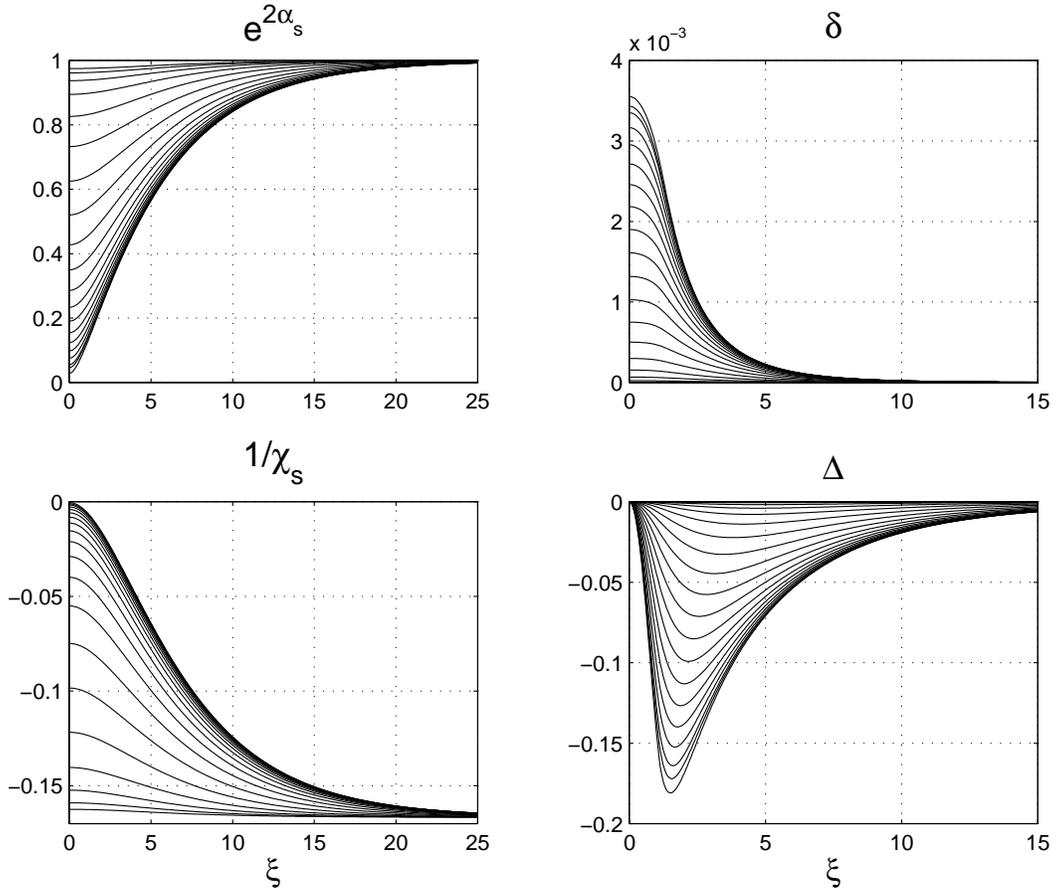,height=12.cm}}} 
%,width=12.cm}}}
\caption{The fields $\alpha_s, \chi_s, \eta,
\Delta_s$ starting with $\xi_0=0, \mu=0.01,
\varepsilon=.2$.}
\label{fig1}
\end{center}
\end{figure}

\begin{figure}[h]
\begin{center}
{\mbox{\epsfig{figure=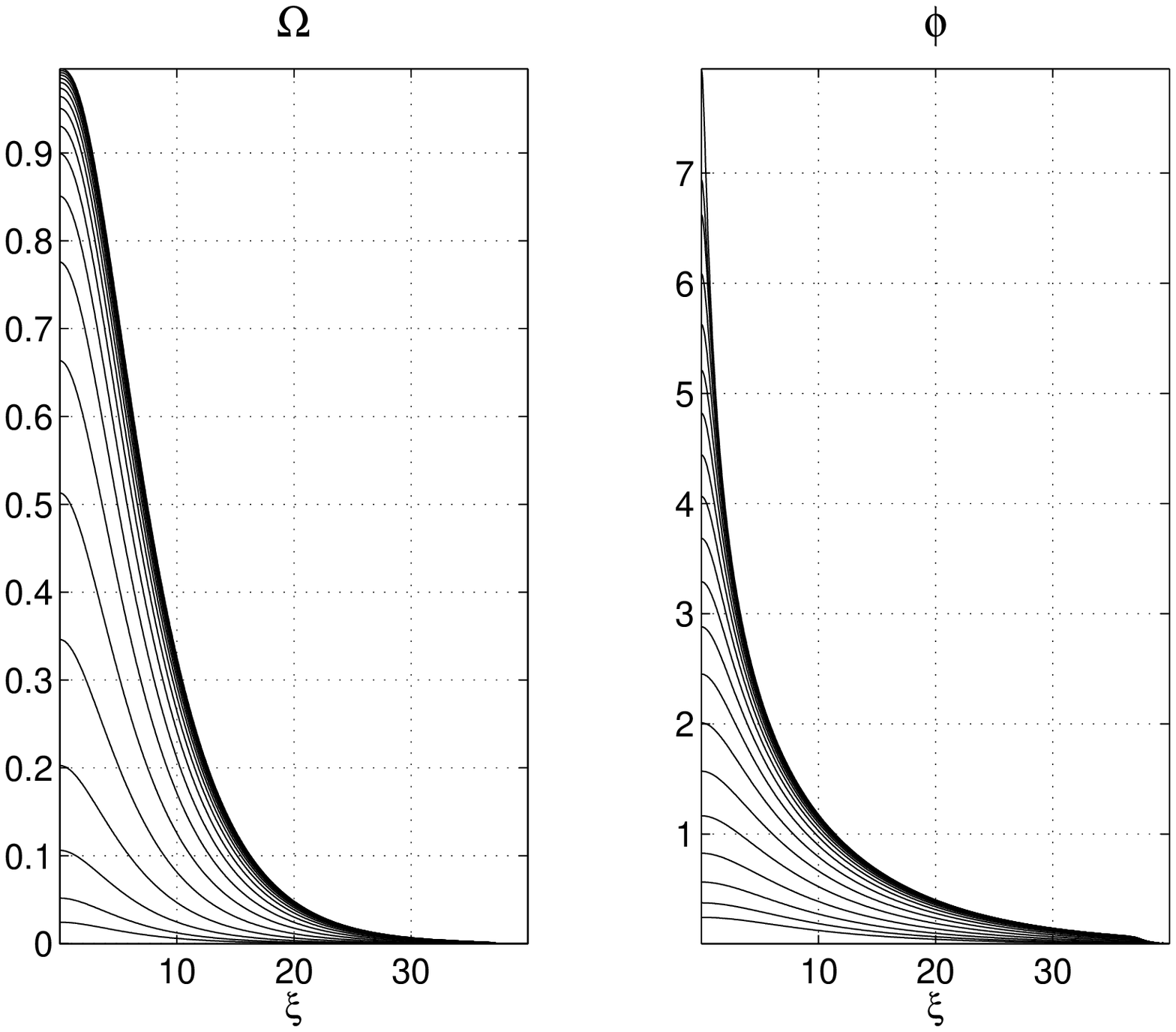,height=8.cm}}} %,width=10.cm}}}
\caption{The $\Omega$ parameter and the dilaton, same initial data as
above.}
\label{fig2}
\end{center}
\end{figure}

\begin{figure}[h]
\begin{center}
{\mbox{\epsfig{figure=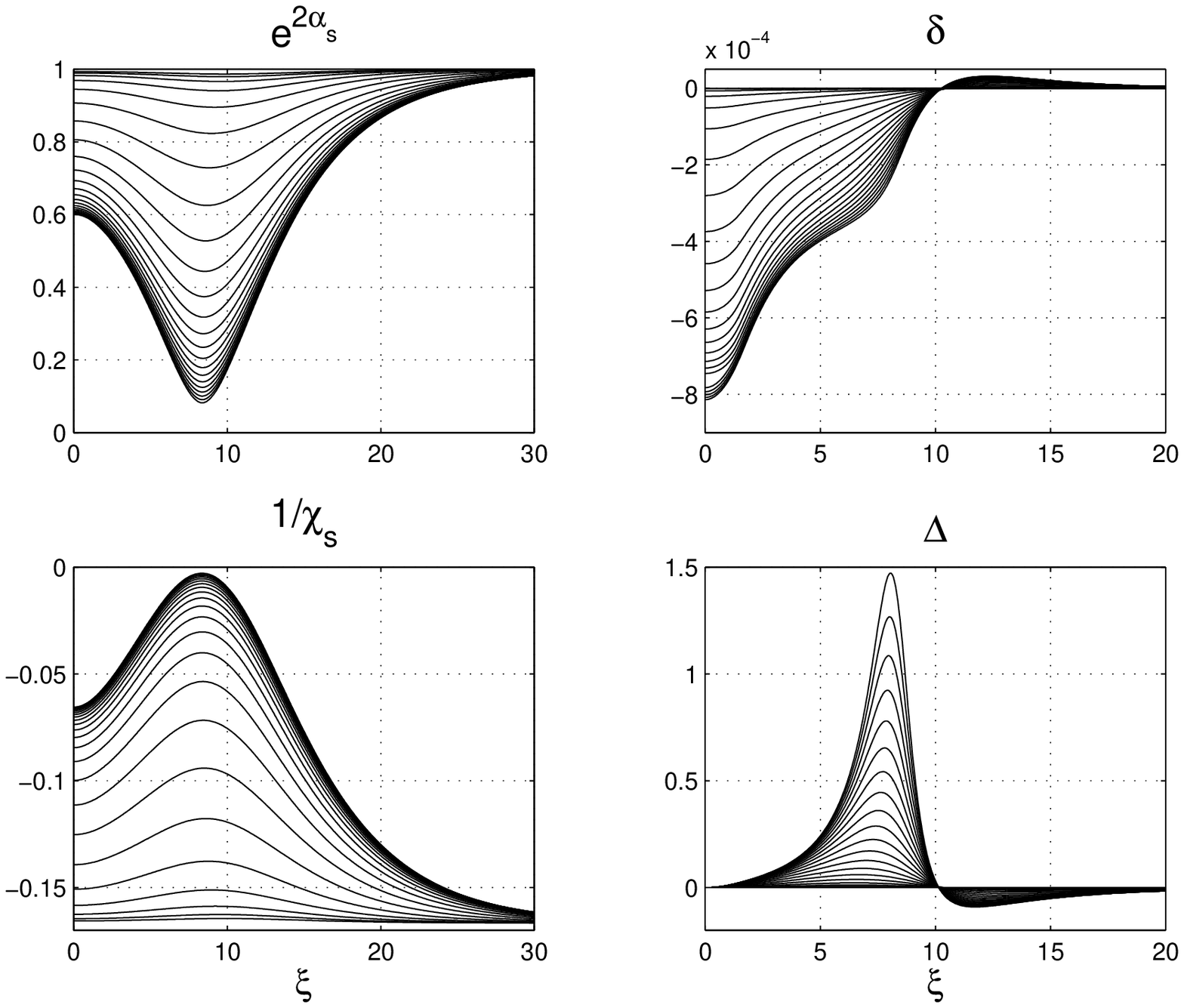,height=12.cm}}} %,width=12.cm}}}
\caption{The fields $\alpha_s, \chi_s, \eta,
\Delta_s$ starting with $\xi_0=10, \mu=0.01,
\varepsilon=0.2$.}
\label{fig1b}
\end{center}
\end{figure}

\begin{figure}[h]
\begin{center}
{\mbox{\epsfig{figure=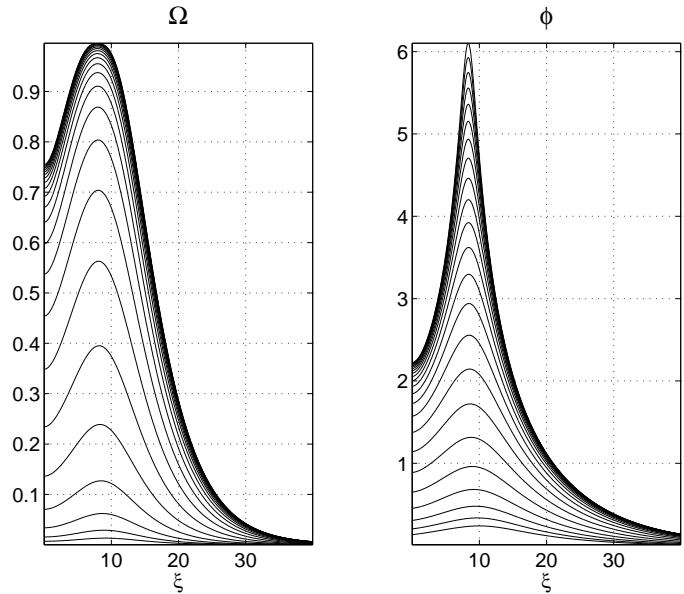,height=8.cm}}} 
%,width=10.cm}}}
\caption{The $\Omega$ parameter and the dilaton, same initial data as
above.}
\label{fig2b}
\end{center}
\end{figure}

%%%%%%%%%%%%%%%%%%%%%%%%%%%%%%%%%%%%%%%%%
We report some preliminary results obtained by running our code with
the initial ansatz given by Eq.\ (\ref{ansatz}) and for several values
of the parameters ($\xi_0, \varepsilon, \mu$); a typical run involves
a grid with $N=4096$ points and takes less than an hour on a modern
workstation.  We set a finite-size $L \approx 10/\varepsilon+\xi_0$
and an ultraviolet cutoff at half way to the size of the Brillouin
zone. The routine {\tt ode45} of {\sc matlab} can be used with its
standard setup; a special care must be devoted to the choice of the
{\sl absolute tolerance} parameter which should be set differently for
the various fields which have very different scales. The evolution
starts at negative $t$ from the ansatz of Eq.\ (\ref{ansatz}); we
report both cases ($\xi_0=0$ and  $\xi_0>0$), which indeed show a
similar pattern (see Figs.(\ref{fig1},\ref{fig1b})).  In
Figs.(\ref{fig2},\ref{fig2b}) both $\Omega$ and the dilaton are
shown.

%%

%%

\begin{figure}[h]
\begin{center}
{\mbox{\epsfig{figure=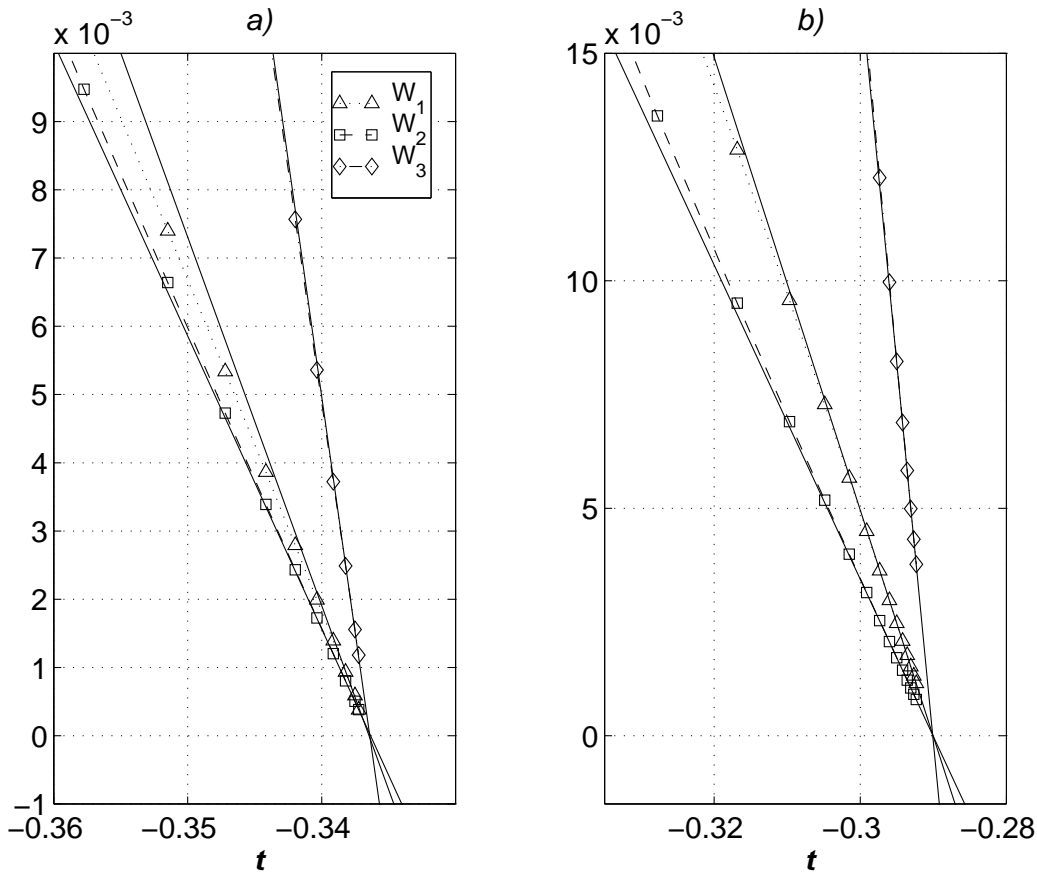,height=12.cm}}} 
%,width=10.cm}}}
\caption{The asymptotic behaviour ${\cal O}(t_0(\xi)-t)$ for the
variables $W_n$ at the $\Omega$ peak ({\it a)} $\xi_0=0$,
{\it b)} $\xi_0=10$)}
\label{fig4}
\end{center}
\end{figure}

We examined the numerical solution in order to check the validity of
the approximation introduced in Sec.\ref{asympt}. In particular we
were able to check the asymptotic behaviour in the region where
$\Omega$ is very close to 1, where Eqs.\ (\ref{asy1}) and (\ref{asy2})
predict a linear regime in time for the following expressions
\begin{eqnarray}
W_1&=&1/\chi\\\nonumber
W_2&=&\exp\{\alpha+2\beta\}\\
W_3&=&\exp\{-(\thrhaf \phi+\alpha+2\beta)/(\surd3-1)\}\nonumber
\end{eqnarray}
(notice that the value of  $\lambda$ is negligible).
This is clearly displayed in Fig.\ref{fig4}.

%\begin{figure}[h]
%\begin{center}
%{\mbox{\epsfig{figure=fig4_0.eps,height=6.cm}}} %,width=10.cm}}}

%\caption{The asymptotic behaviour ${\cal O}(t_0(\xi)-t)$ for the
%variables $W_n$ at the $\Omega$ peak ($\xi_0=0$)}.
%\label{asy0}
%\end{center}
%\end{figure}

%\begin{figure}[h]
%\begin{center}
%{\mbox{\epsfig{figure=fig4_10.eps,height=6.cm}}} 
%%,width=10.cm}}}
%\caption{The asymptotic behaviour ${\cal O}(t_0(\xi)-t)$ for the
%variables $W_n$ at the $\Omega$ peak ($\xi_0=10$).}
%\label{asy10}
%\end{center}
%\end{figure}

\begin{figure}[h]
\begin{center}
{\mbox{\epsfig{figure=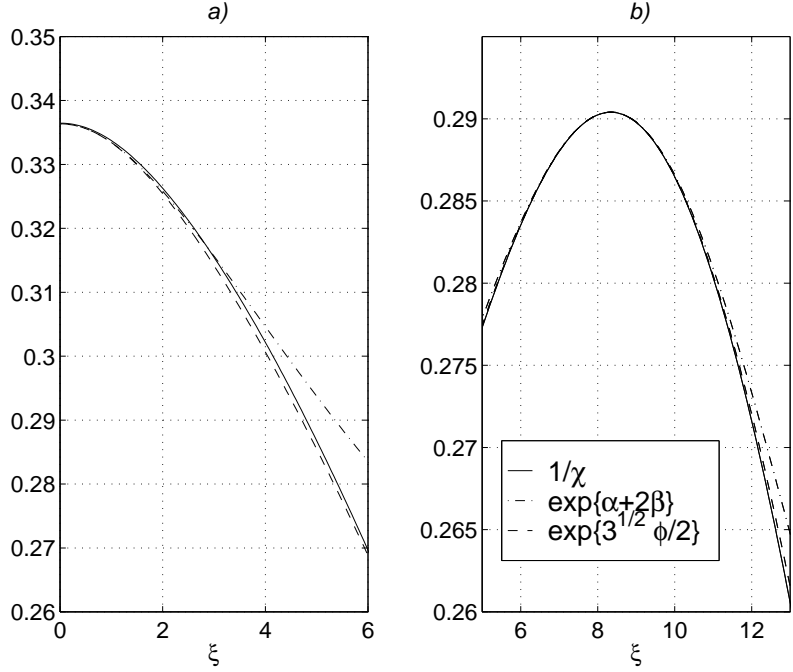,height=9.cm}}} 
%,width=10.cm}}}
\caption{The blowing--up time $t_0(\xi)$, ({\it a)} $\xi_0=0$,
{\it b)} $\xi_0=10$) }.
\label{t00}
\end{center}
\end{figure}

\noindent
From a linear fit to the data, it is then possible to extract
information about the unknown functions $t_0(\xi)$ and $\lambda(\xi)$.
We find that, as expected, $t_0(\xi)$ has an extremum near the
$\Omega$ peak with a small curvature $t(\xi)''$, which is responsible
for the small deviations from the asymptotic estimates of
Eq.\ (\ref{asy1}) (see Fig.\ref{t00}).

\begin{figure}[h]
\begin{center}
{\mbox{\epsfig{figure=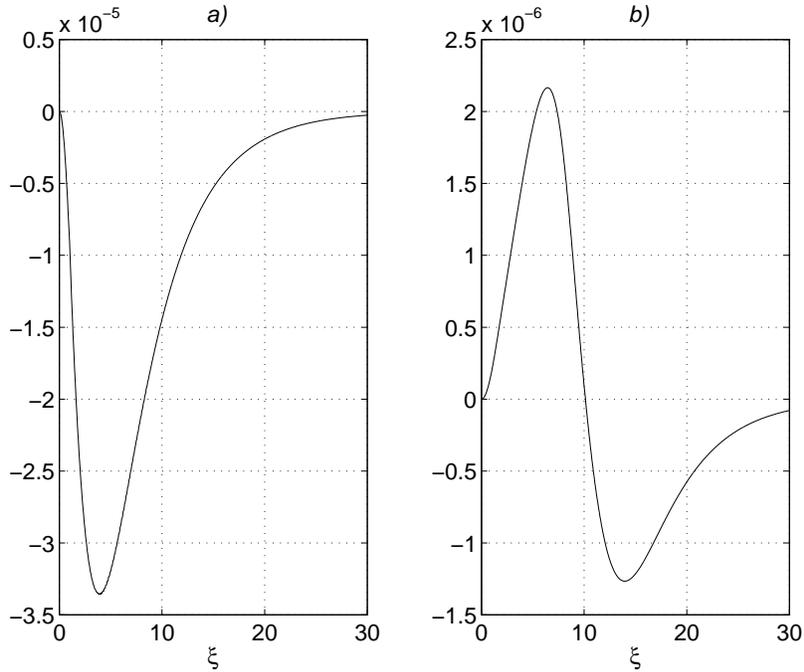,height=9.cm}}}
\caption{The function $\lambda(\xi)$, ({\it a)} $\xi_0=0$,
{\it b)} $\xi_0=10$).}
\label{l00}
\end{center}
\end{figure}

 In all the cases we have examined so far, the function $\lambda(\xi)$
turns out to be very small (see Fig.\ref{l00}).  but this is
probably due to our initial starting point with $\Delta\equiv 0$. We
are currently exploring a wider class of initial data, in order to
find the generic properties of the solutions.

%\begin{figure}[h]
%\begin{center}
%{\mbox{\epsfig{figure=fig5_0.eps,height=6.cm}}} %,width=10.cm}}}

%\caption{The blowing--up time $t_0(\xi)$ ($\xi_0=0$).}
%\label{t00}
%\end{center}
%\end{figure}

%\begin{figure}[h]
%\begin{center}
%{\mbox{\epsfig{figure=fig6_0.eps,height=6.cm}}} %,width=10.cm}}}

%\caption{The function $\lambda(\xi)$ for $\xi_0=0$.}
%\label{l00}
%\end{center}
%\end{figure}

%\begin{figure}[h]
%\begin{center}
%{\mbox{\epsfig{figure=fig6_10.eps,height=6.cm}}} 
%%,width=10.cm}}}
%\caption{The function $\lambda(\xi)$ for $\xi_0=10$.}
%\label{l010}
%\end{center}
%\end{figure}

Finally, our code can also be run backwards i.e. towards $t = -
\infty$. In this case, however, we encounter a problem: although
$\phi$ keeps decreasing in magnitude as $t$ becomes more and more
negative, it also starts to oscillate, as expected \cite{BMUV1}. As a
result, the constraints cannot be always solved with the sign
determination given in (\ref{solconstr}). We have been able to solve
this problem by imposing the constraints only on the initial data
and by otherwise working in an enlarged phase space containing also
$\phi$ and $\dot{\phi}$.  This procedure has the further advantage of
allowing a non-trivial  check of the numerical precision by verifying
that the constraints are conserved.

We expect to present results obtained by following this better
procedure in the near future.

\section{Discussion}
%%%%%%%%%%%%%%%%%%%%%%%%%%%%%%%%%%%%%%%%
In order to discuss better the physical meaning of our result it would
be convenient to transform them back to the original string frame. 
It is difficult to make the change of frame numerically and, indeed,
we think it would be easier to start directly the whole exercise in
the new frame.  Fortunately, we can achieve a semi-quantitative
understanding of the string frame solutions by noticing that:
\begin{itemize}
\item[\sl i)] initially the two frames coincide, since the dilaton 
is constant;
\item[\sl ii)] our numerical results, as the singularity is
approached, fit very well with the analytic asymptotic formulae given
in Section 3.
\end{itemize}
The latter observation allows us to perform the passage to the string
frame analytically (and of course approximately) near the 
singularity.  Following Ref. \cite{GV1} we find rather easily that the
asymptotic metric and dilaton in the string-frame synchronous gauge
are given by
\begin{eqnarray}
ds^2 &=& - dt^2 +  \left[\left(t/t_0 -1
\right)^{2 \alpha_{\xi}}\,
e^{2\tilde{\gamma}} \,d\xi^2 + \left(t/t_0 -1 
\right)^{2\alpha_{\theta,\varphi}}
\,e^{2\tilde{\delta}} d\omega^2 \right] ~~, \nonumber \\
\phi(\xi, t) &=& \phi_0(\xi) + {\gamma_{\phi}}\,
\ln \left({t / t_0(\xi)} -1\right) \; ,\; t \rightarrow t_0
\end{eqnarray}
where, for notational simplicity, we kept denoting by $t$ the string
frame time and by $t_0(\xi)$ the location of the singularity. The
important exponents $\alpha_i$ and $\gamma_{\phi}$ are completely
determined by the function $\lambda(\xi)$ appearing in
Eq.\ (\ref{asy1}). We find:
\begin{eqnarray}
\alpha_{\xi} &=& - {1 \over \sqrt{3}} { 2 \sqrt{1-\lambda^2}
- \sqrt{3} \lambda^2 - 2 \lambda~ (\sqrt{3} + \sqrt{1-\lambda^2})
\over 2 + \lambda^2}~~, \nonumber \\
\alpha_{\theta,\varphi} &=& - {1 \over \sqrt{3}} { 2
\sqrt{1-\lambda^2}
- \sqrt{3} \lambda^2 + \lambda~ (\sqrt{3} + \sqrt{1-\lambda^2})
\over 2 + \lambda^2}~~, \nonumber \\
\gamma_{\phi} &=& -  { 2 (1-\lambda^2)
+ 2 \sqrt{3}  \sqrt{1-\lambda^2}
\over 2 + \lambda^2}~~.
\end{eqnarray}
It is reassuring to check that the $\alpha_i$ satisfy Kasner's
condition:
\begin{equation}
\sum \alpha_i^2 = \alpha_{\xi}^2 + 2 \alpha_{\theta,\varphi}^2 = 1
\end{equation}
and that, furthermore,
\begin{equation}
\gamma_{\phi} -  \sum \alpha_i = -1
\end{equation}
as in the homogeneous case. Also note that, in the limit $\lambda \ll
1$, $ \alpha_i \rightarrow - 1/\sqrt{3}$. This corresponds to the
isotropic limit.  From the above formulae we see that, in the 
string
frame, the metric is typically (super)inflationary since the
inequalities $-1 < \alpha_i < 0$ are almost always fulfilled.  It is
easy to check that, as a consequence, the string frame asymptotic
solution is reliable even if one does not choose time slices
corresponding to $t_0 \sim \rm{constant}$.  The expansion rate is
maximal for the regions where $\lambda$ is very close to zero,
i.e. for very isotropic regions.

We conclude that, as anticipated in refs. \cite{GV1},\cite{BMUV1},
regions with the correct dilaton perturbation do undergo a
superinflationary expansion, become asymptotically flat and, most
likely, isotropic.  It would of course be most interesting to attempt
to generalize our computations to the case of several lumps and/or
shells located randomly in space and with random initial parameters
and see what kind of chaotic inflationary Universe will result.  This,
unfortunately, appears to demand going beyond the spherically
symmetric situation we have considered in this paper.

In conclusion, our numerical approach appears to confirm that, at
least in the case of spherical symmetry, dilaton-driven inflation
naturally  emerges as a classical gravitational instability of small
perturbations around the trivial (Milne) vacuum.  Nonetheless, before
being able to address/answer completely the criticism expressed in
\cite{TW}, the present work needs to be expanded in at least two 
directions: i) classical phase space has to be swept more
systematically, in  particular away from the case of spherical
symmetry; ii) the behaviour at very  early times has to be thoroughly
investigated so that the possibility (see N.  Kaloper et
al. \cite{TW}) that quantum fluctuations at very early times may
destroy the homogeneity needed  for turning on dilaton-driven
inflation is properly assessed. As discussed at the end of Section
V, this latter problem requires a new way of enforcing the constraints
on which we are presently  working and hope to report soon.

\acknowledgements We thank Mr. F.\ Fiaccadori and Dr. F.\ Piazza for
useful collaboration in developing part of the numerical code. One of
us (GV) is grateful to E. Rabinovici for useful discussions concerning
marginal operators and moduli space in CFTs.

\appendix
\section{A procedure to generate a class of admissible initial data}
In this appendix we illustrate our procedure
for choosing the small initial perturbation of Milne's metric
in such a way as to fulfill the necessary  positivity
constraints as well as  regularity at $\xi=0$.
The physical idea is quite simple: since we know that
Milne's solution is regular and satisfies all the constraints, we
construct initial data from a small deformation of Milne.
Notice that there is no need that the deformation
satisfies the evolution equations.

Consider the following deformation of Milne's metric:
\begin{eqnarray}
\alpha &=& {1 \over 4} \ln \left(t^2 - {{ T_0^2} \over {\cosh
{\varepsilon
(\xi - \xi_0)}}} \right) + (\xi \rightarrow - \xi)\; , \nonumber \\
\beta &=& \alpha + \ln~\sinh\xi +\half \zeta \left[{ {T_0^2 \xi^2}
\over {t^2
\cosh {\varepsilon
(\xi - \xi_1)}}} + (\xi \rightarrow - \xi) \right]
\label{deformation}
\end{eqnarray}

Such a deformation is localized at $\xi = \xi_0$, with some anisotropy
concentrated at $\xi_1$. The deformation rapidly drops to zero far
from these two values of $\xi$. The symmetrization is needed in order
to bring the ansatz in the form discussed ion Section 3.  We now
define the initial data from Eqs.\ (\ref{deformation}) and their first
time derivatives at the initial time $t_i \ll T_0$.

We have performed the computation of the initial data using
Mathematica
and checked that the positivity and regularity constraints are
identically satisfied in $\xi$ for a wide choice of the parameters
$t_i/T_0,~\varepsilon,~\xi_0,~\xi_1,~\zeta$, e.g. for the sets used
in section 5:
\begin{eqnarray}
t_i/T_0 = 10, \;\varepsilon = 0.2, \;  \xi_0 = 0 ,\; \zeta =
0  ;\nonumber \\
t_i/T_0 = 10 ,\; \varepsilon = 0.2 ,\; \xi_0 = 10 ,\;  \zeta =
 0 \,. \end{eqnarray}
Indeed, positivity is most sensitive to the parameter $\varepsilon$
which cannot be much larger than $0.5$.  In the limit $t_i/T_0
\rightarrow \infty$ the formulae for the initial data simplify
considerably and positivity constraints can be solved in many cases
analytically.

\begin{figure}[h]
\begin{center}
{\mbox{\epsfig{figure=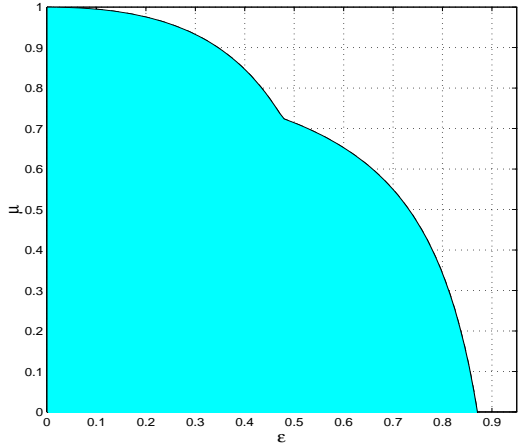,height=6.cm,width=8.cm}}}
\caption{Admissible values for $\mu,\varepsilon$.}
\label{admiss}
\end{center}
\end{figure}

The above initial choice, that we have considered for our numerical
study, is a special case of a class of admissible initial data (we
call a given choice of initial data {\sl admissible} if it satisfies
the constraint $A^2\ge |B|$) which can  be characterised as
follows:

\noindent
{\sc theorem}:{\sl\  Let $f(\xi)$ be twice continuously differentiable
with}
\begin{enumerate}
\item $f(\xi)\ge  f_0 > 1$;
\item $f'(0)=0$;
\item $f''(\xi) \le \varepsilon^2\,f(\xi)$ for some $\varepsilon<1$.
\end{enumerate}
{\sl Then the initial data given by}
\begin{equation}
\alpha_s = \half \ln(1-f(\xi)^{-1})\,,\;
\chi_s=-6\,e^{-2\alpha_s}\,,\; \eta =\,\Sigma\,=0
\;.\label{general}
\end{equation}
{\sl are admissible if}
\begin{equation}
f_0^{-1} \le \max[ {5-4\,\varepsilon-2\,\varepsilon^2\over
6-4\,\varepsilon-2\,\varepsilon^2}\,,\,
1-({2\,\varepsilon \over
3-2\,\varepsilon-\varepsilon^2})^2\;]\;.
\end{equation}
\noindent
We shall first prove the following
\vskip .15in
\noindent
{\sc lemma}: {\sl Let $f(\xi)$ satisfy the conditions of the previous
theorem. Then}
\begin{equation}
-\varepsilon \le {f'(\xi) \over f(\xi)}
 \le \varepsilon\tanh\varepsilon\xi
\end{equation}
\noindent
{\small\sc proof:} Let $g=f'/f$; then it holds
\begin{equation}\label{ineq1}
{dg(\xi)\over d\xi} \le \varepsilon^2 - g(\xi)^2\;.
\end{equation}
First of all let us show that if $g(\overline\xi) < -\varepsilon$
for some $\overline\xi$,
$g$ would develop a singularity at a finite $\xi$: under the 
assumption
and the previous inequality Eq.\ (\ref{ineq1}) it would follow
\begin{equation}
{g'(\xi)\over \varepsilon^2-g(\xi)^2} \ge 1 \nonumber
\end{equation}
which can be integrated to give
\begin{equation}
\int_{g(\xi)}^{g(\overline\xi)} {dg \over g^2-\varepsilon^2}
\ge \xi - \overline\xi \;.\nonumber
\end{equation}
Since the integral is convergent, $g\to-\infty$ at a finite $\xi$;
$f$ being regular and $>1$ this is a contradiction.
Assume next that $g(\xi)>\varepsilon\tanh\varepsilon\xi$ at some
$\overline\xi$; by continuity we may assume
$g(\overline\xi)<\varepsilon$. It follows
\begin{equation}
{g'(\xi)\over \varepsilon^2-g(\xi)^2} \le 1\;;\nonumber
\end{equation}
by integrating from $\xi=0$ we find
$\tanh^{-1}(g(\overline\xi)/\varepsilon)\le 
\varepsilon\overline\xi$,
 a contradiction. This completes the proof of the lemma.
\vskip .15in
\noindent
To prove the main theorem, let us insert Eq.\ (\ref{general}) into 
the
expression of $A_s^2 \pm B_s$ (Eq.\ (\ref{ABs})); we get
\begin{eqnarray}
(f-1)\,\exp\{2\alpha_s\}\,\left(A_s^2 \pm B_s\right) &=& \\
3-{f''\over f} &+& {2f - 5/4 \over f-1}\,\left({f'\over f}\right)^2 -
2\, {f'\over f}\, \coth\xi \,\pm \,2\,{f'\over f}\sqrt{f\over 
f-1}\;.\nonumber
\end{eqnarray}
We can apply the lemma at once and see that the l.h.s. is bound from
below as follows
\begin{equation}
{\rm l.h.s.} \ge 3-\varepsilon^2-2\,\varepsilon + 2 \left({f'\over
f}\right)^2 \,
\pm \,2\,{f'\over f}\,
\sqrt{f\over f-1}\nonumber
\end{equation}
or, we may complete the square before applying the inequality,
\begin{equation}
{\rm l.h.s.} \ge 3-  \varepsilon^2-2\,\varepsilon -\half {f\over
f-1} + 2\left({f'\over f}\pm \half \sqrt{{f\over 
f-1}}\right)^2\nonumber
\end{equation}
and the result follows. Notice that the theorem covers the cases that
we have considered in section \ref{numerical}; the domain of 
parameters for which the data are admissible is shown in
Fig.\ref{admiss}.

The class of initial data which has been defined here depends on a
single arbitrary function. The general expression of $f$ can be easily
derived by setting $\varepsilon^2 f - f'' = \rho(\xi) \ge 0$. The
solution is clearly
\begin{equation}
f(\xi) = \mu^{-1}\cosh(\varepsilon\xi)
- \int_0^\xi \varepsilon^{-1}\sinh\left(\varepsilon(\xi-\xi')\right)
\,\rho(\xi')\,d\xi'\;.
\end{equation}
If $\int_0^\infty \rho(\xi)\,d\xi\,$ is small enough, then we get
admissible initial data. As stressed in Sec.\ref{fieldeq}, it would be
interesting to have a general characterisation of suitable Cauchy 
data.

\end{document}